\documentclass[journal]{IEEEtran}  
\usepackage{lipsum}
\usepackage{amsmath}
\usepackage{graphicx}
\usepackage[export]{adjustbox}
\usepackage{amssymb}
\usepackage[caption=false]{subfig}
\usepackage{footnote}
\usepackage[noend]{algpseudocode}
\usepackage[utf8]{inputenc}
\graphicspath{{images/}}
\usepackage{cite}
\usepackage{optidef}
\usepackage{amsthm}
\usepackage[ruled,linesnumbered]{algorithm2e}

\usepackage{booktabs}
\usepackage{multirow}

\title{Context-Aware Wireless Connectivity and Processing Unit Optimization for IoT Networks}

\author{\IEEEauthorblockN{Metin~Ozturk\IEEEauthorrefmark{1}, Attai~Ibrahim~Abubakar\IEEEauthorrefmark{1}, Rao~Naveed~Bin~Rais\IEEEauthorrefmark{2}, Mona Jaber\IEEEauthorrefmark{3}, Sajjad~Hussain\IEEEauthorrefmark{1},  Muhammad~Ali~Imran\IEEEauthorrefmark{1} }\\
\thanks{{\IEEEauthorrefmark{1}Communication, sensing and imaging (CSI) research group, 
James Watt School of Engineering, University of Glasgow, United Kingdom. 

\IEEEauthorrefmark{2}Electrical and Computer Engineering, Ajman University, UAE.

\IEEEauthorrefmark{3}School of Electronic Engineering and Computer Science, Queen Mary University of London, United Kingdom.}

This work was supported by EPSRC Global Challenges Research Fund the DARE Project: Grant EP/P028764/1, and partly funded by Ajman University.
The first author was supported by the Republic of Turkey Ministry of National Education.}
}

\begin{document}

\maketitle
\begin{abstract}
A novel approach is presented in this work for context-aware connectivity and processing optimization of Internet of things~(IoT) networks.
Different from the state-of-the-art approaches, the proposed approach simultaneously selects the best connectivity and processing unit~(e.g., device, fog, and cloud) along with the percentage of data to be offloaded by jointly optimizing energy consumption, response-time, security, and monetary cost. 
The proposed scheme employs a reinforcement learning algorithm, and manages to achieve significant gains compared to deterministic solutions.
In particular, the requirements of IoT devices in terms of response-time and security are taken as inputs along with the remaining battery level of the devices, and the developed algorithm returns an optimized policy.
The results obtained show that only our method is able to meet the holistic multi-objective optimisation criteria, albeit, the benchmark approaches may achieve better results on a particular metric at the cost of failing to reach the other targets.
Thus, the proposed approach is a device-centric and context-aware solution that accounts for the monetary and battery constraints. 
\end{abstract}

\section{Introduction}
The world is witnessing a rapid adaptation of the Internet of things~(IoT) technologies to enhance and augment the human endeavor in operating almost every aspect of our lives.
IoT is a system of connected devices that sense and actuate signals in the real world.
These devices are connected and enable real-time data collection and respective optimized action.
However, designing and deployment of IoT networks pose multiple challenges, including energy consumption, connectivity, and data processing~\cite{AkpakwuIoTsurvey,PoplisurvNBIOT,ozturk2018energy}. 

As previously stated in~\cite{KouzayhaMeasIoT}, different communication protocols and standards exist, with each having its pros and cons.
Cellular IoT networks, for example, operate in the licensed band and utilize mobile network infrastructure for IoT device inter-connectivity. 
They are the most suitable option for private data transmission because they provide reliable, secure~(due to the presence of eSIM card), and wide range coverage with vast infrastructure.  
Three cellular technologies have been proposed by 3GPP Rel-13 to support IoT connectivity: narrowband IoT~(NB-IoT), enhanced machine type communications~(eMTC), and enhanced coverage GSM~(EC-GSM)~\cite{PoplisurvNBIOT}.
NB-IoT has been designed to offer low cost, wide coverage, and low data rate transmission with limited mobility support. 
The short-range solution, on the other hand, comprises Zigbee, Wi-Fi, Bluetooth, etc. which could also be publicly, privately or jointly owned~\cite{PalattellaIoT5G}.

As argued in~\cite{smart-port-dct}, the smart port concept is adopted in this research to illustrate context-aware connectivity and processing.
Smart ports represent a vast application as they bridge the fleet of vessels to the fleet of trucks and include various aspects of Industry 4.0 technologies.
Moreover, cost-efficiency is a prime objective in the application of smart ports in view of the competition between various parties in quality and price. 
Towards that end, as the industry looks for least complexity in IoT devices to reduce their capital and operation cost and enable cost-effective applications, the design of an IoT system faces the paradigm of optimizing the location of the data processing.
This could be in the cloud or device but also in an intermediate node (e.g., Wi-Fi gateway) referred to as fog.

Furthermore, the massive deployment of IoT devices would result in the generation of a huge amount of data which would be difficult for traditional processing techniques to handle~\cite{Sun2019AIIOT}.
In addition, due the massive inter-connectivity of devices, network management and orchestration will become very complex and challenging, hence the need for more intelligence to be included in IoT networks to enable enhanced data processing as well as intelligent and autonomous network operation~\cite{MadeepIoT}. 
In the midst of diverse IoT network standards and protocols, vast IoT applications, varying characteristics of IoT devices (stationary, mobile, battery life, etc.), and several connectivity options, there is the need to carefully select the most suitable method to harness, store, transmit, process, and secure the data obtained from massive IoT networks in an energy efficient and cost effective way. 
The application of artificial intelligence~(AI) and machine learning~(ML) would help harness the useful information embedded in the massive data that is generated in order to facilitate decision making while ensuring smart and efficient network operation.

\subsection{Related Work}
A review of  various technologies, protocols and standards that can be utilized for IoT networks was carried out in~\cite{PoplisurvNBIOT, ElsaadanyLTEIOT, PalattellaIoT5G} while in~\cite{HassijaIoTsec2019, haddadpajouh2019survey} a survey of various security challenges and solutions in IoT networks was performed.
The authors in ~\cite{PersiaIOT2019} carried out a comparative study of the connectivity based performance of two IoT enabling technologies, namely NB-IoT and LoRa.
One of the major concerns of IoT networks is the energy consumption of IoT devices, hence the network has to be designed and optimized in such a way that it ensures the longevity of the device battery life.
In this regard, the authors in~\cite{abegundeEEIoT} proposed a smart game algorithm to optimize both data transmission and energy consumption.

The choice of where to process and store data is also one of the decisions that need to made in IoT networks due to the limited processing, storage as well as battery life of IoT devices.
In this regard, different caching and offloading strategies have been developed.
The authors in~\cite{TayadeEC2017, tayade2018delay} considered the trade-off between the energy consumed for local data processing at device and that due to offloading to the edge cloud.
The former proposed a theoretical framework for the reduction of the energy consumption of the devices by developing an optimal offloading strategy for all user devices in the network.
The later considered the delays---due to offloading data to the edge cloud for processing---as an optimization constraint while developing a close form expression for determining the optimal offloading strategy for multiple user devices.
A novel energy and context-centric scheme based on ML was introduced in~\cite{BiasonEC2017}, which aims at elongating the device battery life while ensuring that the required quality of service~(QoS) level is maintained.

Deep learning approaches are also exploited for IoT data analytic and network optimization. 
A deep learning model was proposed in~\cite{Lee2019} for joint transmission and recognition for IoT devices in order to ensure efficient data transmission to the server for recognition purposes under very low signal-to-noise-ratio~(SNR). 
In~\cite{Lee2019EE}, a deep learning based framework for forecasting the energy consumption of IoT networks in an edge computing scenarios was introduced.
A novel offloading strategy based on deep learning was developed in~\cite{Li2018IOT} to enhance the performance of IoT networks as well as optimize the amount of edge computing tasks that can be performed.
Another important aspect of IoT networks is ensuring the confidentiality of the massive data that is collected from IoT devices.
To that end, the authors in~\cite{Guo2019privacy} proposed a deep learning framework for big data analytic in IoT networks that would ensure that the privacy of the data is preserved. 

However, most of the proposed approaches considered a layered approach to IoT network optimization, where one or two optimization constraints were considered among the many of which include energy consumption, connectivity, transmission delays, storage, computing/processing, and monetary considerations, respectively.
We argue that a holistic solution approach that considers several optimization constraints is necessary for IoT network optimization in order to fully harness the benefits that such solution has to offer.
In this regard, the authors in~\cite{ozturk2018energy} proposed a context-aware scheme for IoT networks for the joint optimization of computation and connectivity using $Q$-learning, an reinforcement learning~(RL) algorithm. 
A two stage $Q$-learning algorithm was developed where the first stage involved the selection of the connection-processing unit pair, while the second stage was to determine the amount of data to be offloaded to fog or cloud. 
However, the penalty function developed for the $Q$-learning algorithm is not scalable, and hence difficult to adapt it to changing IoT application requirements.
Moreover, in the proposed approach, it is not possible to prioritize the importance of one constraint over another, which helps enhance the context-awareness of the solution.

\subsection{Objectives and contributions}
RL is employed in this work in order to manage multiple optimization objectives, such as connectivity, processing, storage, etc., jointly.
The following objectives are pursued in this work:
\begin{itemize}
    \item \textbf{energy consumption and monetary cost:} based on the battery conditions of IoT devices, the total energy consumption of a device---incurred by data processing and transmission---along with monetary cost is minimized.
    \item \textbf{quality requirements of IoT devices:} meeting the stipulated quality of the IoT device/application is another objective of this optimization problem.
    In this work, we focus on security and latency exigencies.
\end{itemize}

In that regard, in this work, a context-aware connectivity and processing optimization in IoT networks using $Q$-learning is developed for the joint optimization of connection-type, processing unit, percentage of data to be offloaded, response time, security as well as monetary cost.
Different from~\cite{ozturk2018energy}, the proposed holistic framework for IoT network optimization involves only one stage of $Q$-learning. 
A simpler and more generalized penalty function is also introduced to capture the variations in IoT application demands. 
In addition, a weighing mechanism is designed that will enable the IoT devices to prioritize any of the optimization constraints, on which basis, the algorithm would change its behaviour.  

The remaining parts of this paper is organised as follows: The system model is defined in Section~\ref{sec:system-model}, while the considered problem is formulated in Section~\ref{sec:problem-formulation}. 
The proposed RL based IoT optimization framework is presented in Section~\ref{sec:QL}, and the performance evaluation via a comprehensive results analysis are discussed in Section~\ref{sec:performance}.
Lastly, Section~\ref{sec:conclusion} concludes the article. 

\section{System Model}\label{sec:system-model}
\begin{figure}[h]
	\centering
	\includegraphics[width=0.5\textwidth]{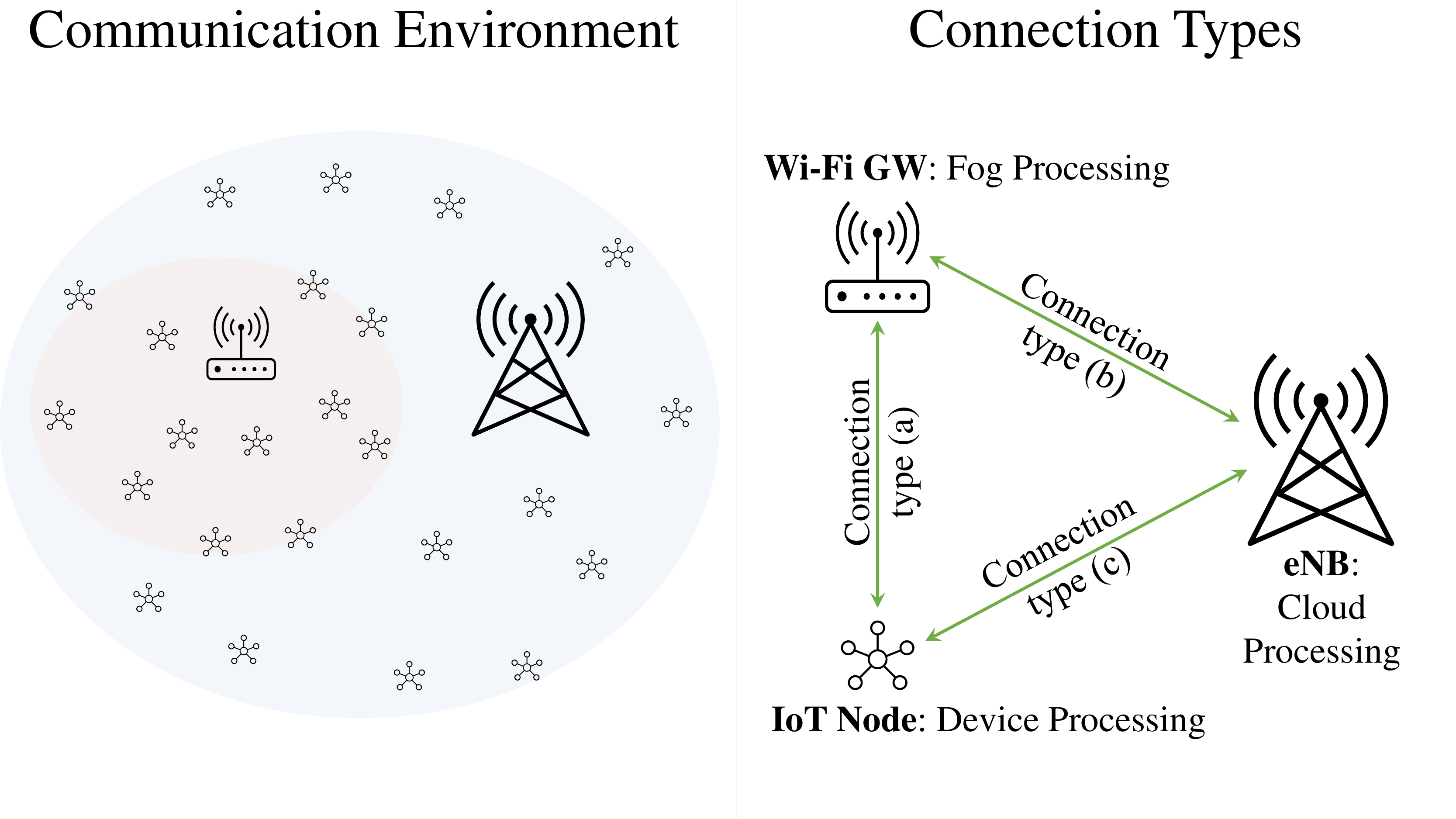}
	\caption{System modelling. The IoT network scenario is depicted on the left hand side, while the descriptions of the considered wireless connection types. While the eNB has a fixed location throughout all the repeating simulations, the gateways and IoT devices are located randomly, and thus their locations may vary from this sample snapshot. GW: gateway}\label{fig:simuli}
\end{figure}

A novel scheme for context-aware connectivity and processing developed in this work may be applied to any IoT application. 
For clarity purposes, the system model is built based on the smart port environment as depicted in Fig.~\ref{fig:simuli}.
The IoT devices are powered by batteries of varying capacity and lifespan.
They are all equipped with a measure of processing ability required for basic operation and are able to offload their data processing operations to Wi-Fi gateway~(fog) or the base station (cloud).

\subsection{Propagation Model}
As seen in Fig.~\ref{fig:simuli}, the following wireless connections are modelled: (a) Device-to-Gateway (Wi-Fi), (b) Device-to-eNB (NB-IoT), and (c) Gateway-to-eNB (LTE).
On the one hand, as we considered that there are neighbouring Wi-Fi gateways operating at the same frequency---which is a quite likely scenario---, connection type (a) is taken to be interference limited.
On the other hand, it is assumed that i) NB-IoT technology is not used in other surrounding eNBs and ii) a scheduler is used for LTE connections, thereby connection types (b) and (c) are considered to be noise limited.

The propagation modelling is performed here in order to obtain the amount of required transmit power for all the connection types, which is subsequently used for energy consumption computations.
In general, wireless communication is governed by propagation loss that is a function of distance and frequency.
We adopt the log-distance path-loss~($L$) model as follows:
\begin{subequations}
	\begin{eqnarray}
	L=L_0+10\delta \log_{10}\dfrac{d}{d_0} + \xi_\text{g}, \\
	L_0 = 20\log_{10}\left(\dfrac{4\pi d_0}{\lambda}\right),
	\end{eqnarray}
\end{subequations}
where $\delta$ is the path-loss exponent, $d$~(in meter) is the distance between the transmitter and receiver with $d_0$~(in meter) being the reference distance.
$\xi_\text{g}$ is the shadowing component with a standard deviation, $\sigma$, and zero mean.
$L_0$ is the path-loss at the reference distance $d_0$, $\lambda=\dfrac{c}{f_\text{c}}$ is the wavelength, where $c$ is the speed of light, and $f_\text{c}$ is the transmission frequency. 

Signal-to-interference-plus-noise ratio~(SINR) is an important measure, since it determines the level of signal power at the receiver, which imposes a certain level of received signal power---referred as receiver sensitivity, under which the communication link fails.
Moreover, the receiver sensitivity varies for different radio access technologies~(RATs), thereby the communication channel should be designed accordingly.
Given that various communication types~(e.g., LTE, NB-IoT, and Wi-Fi) with diverse receiver sensitivities are considered in this work, the link margin concept is used to capture the distinctive sensitivity levels.

From Shannon's channel capacity theorem, we know that the achievable data rate is a function of SINR, such that higher SINR values result in higher data rates, and vice versa.
As such, by manipulating the well-known Shannon capacity formula, we obtain the receivable data rate as follows~\cite{TayadeEC2017,ozturk2018energy}:
\begin{equation}\label{eq:D}
D=T B\log_2 \left(1+\frac{P_\text{r}}{P_\text{I}+\mathcal{N}_0 B}\right),
\end{equation}
where $P_\text{r}$ is the required received power, $D$ (in bits) is the finite-length data to be transmitted, $T$ is the time period, $B$ is the channel bandwidth, and $P_\text{I}$ is the cumulative interference power on the given channel during time period $T$.
Please note that $P_\text{I}$ has no effect for wireless connections of type (b) and (c), since they are noise limited.
Next, the required received power---which is needed to achieve the target data transmission---is calculated using~\eqref{eq:D} and solving for~$P_\text{r}$~\cite{TayadeEC2017,ozturk2018energy}:
\begin{equation}\label{eq:Pt}
P_\text{r}=\left(2^{\frac{D}{T B}}-1\right) (P_\text{I}+\mathcal{N}_0B).
\end{equation}
In this regard, the IoT devices are aware of their data rate requirements, which is then used to compute the required transmit power through~\eqref{eq:Pt}.

\subsection{Energy consumption model}
Similar to~\cite{ozturk2018energy}, two main components of the energy consumption is considered and modelled: wireless transmission and task computation.
The energy consumption due to wireless transmission is represented by $E_\text{tx}$, while that due to task computation is denoted by $E_\text{p}$, hence the total energy consumption in the IoT network is obtained by summing both together.
The energy as a result of transmission power is dependent on the transmission route selected by the IoT device. 
The transmission route can be either one hop from device to NB-IoT~($E_\text{tx,b}$) or two hops with the first hop being device to Wi-Fi gateway while the second hop Wi-Fi gateway to LTE~($E_\text{tx,a} + E_\text{tx,c}$).
The energy consumption due to task processing is determined by the amount of data~($D$), computing capability of the processing unit\footnote{It is measured in the number of computational cycle per data element~(DE); i.e., higher $X$ yields less computational power} ($X$) as well as the energy consumed in each computation cycle~($\epsilon$); thus $E_\text{p} = f(D,X, \epsilon) = \frac{D}{W}X\epsilon$, where $W$ is the number of bits per DE~\cite{TayadeEC2017}.

\subsection{Response time model}\label{sec:response-time-model}
The response time is modelled in this work in a similar fashion to~\cite{ozturk2018energy}.
The response time experienced by an IoT device is a function of the latency experienced during transmission from IoT device to the server.
We model the uplink delay in this paper, while assuming that all devices have the same downlink delay. 
Towards that end, task computation~(processing delay, $t_\text{p}$) and data transmission~(transmission delay, $t_\text{t}$) are two primary elements of the uplink delay.
Of these two, the processing delay~($t_\text{p}$) is a function of the processor's computational power~($X$), such that more delay is experienced with a lower computational power at the processor.
A server usually possesses much greater processing power compared to an IoT device, with gateway in between, such that $X_\text{c} < X_\text{f} < X_\text{d}$, where $X_\text{d}, X_\text{f}, X_\text{c}$ are the computational powers of device, fog, and cloud, respectively.
Hence, we model $t_\text{p}$ with respect to the processing powers of the various location where task computation will be performed as follows: $\frac{t_\text{p,d}}{X_d} = \frac{t_\text{p,f}}{X_\text{f}} = \frac{t_\text{p,c}}{X_c}$, with $t_\text{p,d}$,
$t_\text{p,f}$ and $t_\text{p,c}$ denoting delays due to task processing at the device, fog, and cloud, respectively.
The output from the processing stage is usually a compressed data and is of lesser volume compared to the original raw input data.  
As a result, the compression rate that accounts for the difference in data volume between raw input and processed output data is expressed as $\raisebox{\depth}{\scalebox{-1}[-1]{$\Omega$}}$; $D_\text{r} = \raisebox{\depth}{\scalebox{-1}[-1]{$\Omega$}}D_\text{p}$ with $D_\text{r}$ and $D_\text{p}$ denoting the volume of input~(raw) and output~(processed/compressed) data, respectively.

The type of RAT as well as the amount of data to be transmitted has a major impact on the level of transmission delays that will be experienced. 
The use of Wi-Fi technology causes more transmission delays because it uses the unlicensed frequency band, making it more prone to frequent re-transmission due to frequent collisions.
Hence, in this research, as it is done in~\cite{ozturk2018energy}, we consider this effect by introducing a factor $F>1$, such that delay experienced due to data re-transmission over Wi-Fi is $F$ times greater than that over LTE or NB-IoT; $t_\text{t,a} = t_\text{t,b}F = t_\text{t,c}F$, where $t_\text{t,a}$, $t_\text{t,b}$ and $t_\text{t,c}$ represents the transmission delays due to connection type (a), (b), and (c), respectively.
This is captured in the model in Fig.~\ref{fig:simuli}, whereby the IoT device or the gateway could be the source with either the gateway or cloud being the receiver.

As a result, the total response time per action can be computed as:
\begin{equation}
\label{eq:response-time-model}
    R = t_\text{p}D + \sum_{i = 1}^{N_\text{h}} t_\text{t,i} D_\text{i},
\end{equation}
where $N_\text{h}$ is the number of hops, and $D\in \{D_\text{r},D_\text{p}\}$, while $t_\text{t,i}$ and $D_\text{i}$ denotes the values of $t_\text{i}$ and $D$ for the $i^\text{th}$ hop respectively.
It should be noted that connection type (a) and (b) represent the first hop, while the second hop is depicted by connection type (c).

\section{Problem Formulation}\label{sec:problem-formulation}
\begin{figure}[h]
	\centering
	\includegraphics[width=0.5\textwidth]{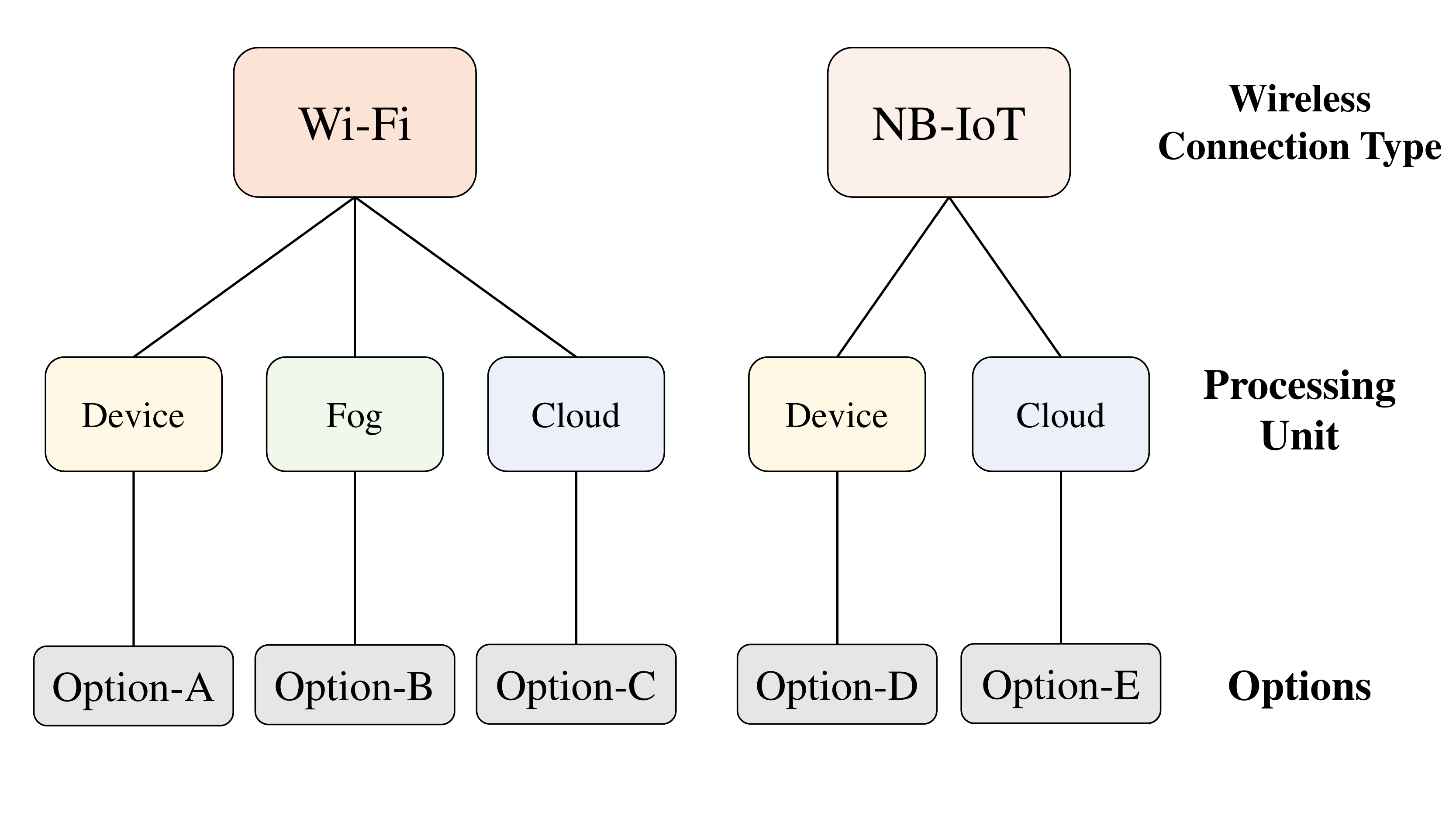}
	\caption{Possible options for a pair of connection type and processing unit. Note that although device, fog, and cloud processing are all available for Wi-Fi case, NB-IoT includes only device and cloud processing owing to the fact that it does not have/require a Wi-Fi gateway to connect to the Internet.}\label{fig:iot-options}
\end{figure}

All the possible connection types and processing unit pairs are given in Fig.~\ref{fig:iot-options}.

Considering the diversity in the available options provided in Fig.~\ref{fig:iot-options}, there are multiple components and objectives of the developed optimization problem:
	\paragraph{Requirements of IoT devices} there could be a broad range of requirements based on the use-case, scenario, and conditions, such as data rate, security, latency, etc., to name a few.
	In this work, response time---as another interpretation of latency---and security are considered as possible requirements of IoT devices.
	As such, let $K_\text{r}$ be the response time requirement of an IoT device, and $\hat{K}_\text{r}$ be the response time offered by the selected option.
	Therefore, in order to satisfy the response time requirement of the IoT device, the following criterion must be met:
	\begin{equation}
		\hat{K}_\text{r} \leq K_\text{r}.
	\end{equation}
	
	The security requirement, on the other hand, is captured by eSIM protection, such that IoT devices opt for eSIM protection if data security is of importance to them.
	In a more formal way, let $K_\text{s}\in\{0,1\}$ be the security requirement of an IoT device, and $\hat{K}_\text{s}\in \{0,1\}$ be the level of data security offered by the selected wireless technology, where 1 indicates the need for eSIM protection and 0 means eSIM protection is unnecessary.
	In this regard, the following condition is needed in order to meet the security requirement of the IoT device:
	\begin{equation}
		\hat{K}_\text{s} \geq K_\text{s}.
	\end{equation}
	
	\paragraph{Energy consumption} the total energy consumption of the IoT device~($E_\mathbb{T}$), which consists of $E_\text{tx}$ and $E_\text{p}$, should be minimized in order to keep the device alive for longer.
	However, the requirements of IoT devices may undermine this objective, since the number of available options may reduce in an attempt to satisfy the requirements.
	Furthermore, the battery level of an IoT device, represented by $\beta$, is also considered in this work provided that it could be another variable in the optimization problem, such that the energy consumption of the device can be prioritized if $\beta$ goes low.
	
	\paragraph{Monetary cost} it is also important to reduce the monetary costs due to the fact that a smart port scenario, which is more related to trading and business, is targeted in this work.
	Since lower costs is vital to keep the business sustainable and profitable, the total data processing cost, represented by $M_\mathbb{T}$, is supposed to be minimized as well, constituting another objective for the optimization problem.
	To this end, each processing unit incurs different costs: $M_\text{d}$, $M_\text{f}$, and $M_\text{c}$ are the processing cost per bit for device, fog, and cloud processing, respectively.

In addition to the requirements and objectives, the optimization problem has strict constraints as well.
For example, the available computational capacity of the processing unit is the major constraint, since it can render the selection infeasible.
In this regard, let $\hat{K}_\text{x} \in \{X_\text{d}, X_\text{f}, X_\text{c}\}$ be the computational capacity of the selected processing unit, where $X_\text{d}$, $X_\text{f}$, and $X_\text{c}$ are the available capacities of device, fog, and cloud, respectively.
Then, the following condition must be obeyed to ensure that the selected processing unit has enough capacity for the required amount of data:
\begin{equation}\label{eq:capacity-constraint}
	K_\text{x} \leq \hat{K}_\text{x},
\end{equation}
where $K_\text{x}$ is the required data rate by the IoT device.
Note that although $K_\text{x}=D_\text{r}$, the $K_\text{x}$ notation is kept here for the sake of consistency.

In addition to the computational capacity, the aforementioned requirements of IoT devices create additional constraints due to the fact that poorly addressed response time and security requirements can make the whole process impractical.
For example, meeting the security requirement can be a must for some use-cases, which value data privacy and cannot tolerate data breaches.
In this regard, for such use-cases, Options A, B, and C in Fig.~\ref{fig:iot-options} are eliminated from the possibilities, since they are with Wi-Fi connectivity that does not offer an eSIM protection.

Moreover, partial offloading---where IoT devices are allowed to offload portions of their raw data to the fog or cloud---is also considered in this work.
As such, the amount of data to be offloaded should also be optimized, thereby we propose a joint optimization of connection-processor pair and amount of data to be offloaded.
The reasoning behind the partial offloading concept is that it is not sufficient to optimize the best connection-processor pair alone, since this kind of optimization cannot be done properly without considering and optimizing the amount of data to be offloaded given the aforementioned constraints.

In this regard, the overall optimization problem can be written formally as follows:
\begin{equation}\label{eq:iot-optimization}
\begin{aligned}
\min_{K, \Psi} \quad & E_\mathbb{T}(K, \psi), M_\mathbb{T}(K, \psi)\\
\textrm{s.t.} \quad & \hat{K}_\text{r} \leq K_\text{r},\\
&\hat{K}_\text{s} \geq K_\text{s}, \\
&K_\text{x} \leq \hat{K}_\text{x},\\
\end{aligned}
\end{equation}
where $\psi$ is the selected percentage volume of data to be offloaded.
$K\in \{\text{A}, \text{B}, \text{C}, \text{D}, \text{E}\}$ is the selected option from Fig.~\ref{fig:iot-options}, and is a 5-tuple as follows:
\begin{equation}
	K=[\hat{K}_\text{r}, \hat{K}_\text{s}, \hat{K}_\text{x}, \hat{E}_{\mathbb{T}},\hat{M}_{\mathbb{T}}],
\end{equation}
where $\hat{E}_{\mathbb{T}}$ and $\hat{M}_{\mathbb{T}}$ are the resulting total energy consumption and monetary cost offered by the selected option, $K$, respectively.

\section{Proposed Scheme}\label{sec:QL}
Given its promising convergence features and capabilities of working in dynamically changing environments, $Q$-learning---one of the most popular RL algorithms---is employed in this work~\cite{qLearningSurvey, qLearning, Qconverge}.
In general, $Q$-learning is an algorithm, where an agent interacts with its environment by taking actions and evaluating consequent outcomes---referred to as reward or penalty.
In particular, there are four main components of $Q$-learning~\cite{qLearningSurvey, qLearning}. 1) \textit{environment:} anything that produces an output for any taken action; 2) \textit{agent:} the entity that takes the actions within the given environment; 3) \textit{action:} the movements that the agent performs to observe the output; 4) \textit{state:} the condition of the agent according to the action taken.

It is a model-free algorithm, meaning that it does not require a prior knowledge about the environment; it rather learns the environment through continuous interactions.
In addition to the aforementioned convergence and adaptability features, this model-free characteristics of $Q$-learning also makes it a good candidate for the problem defined and modelled in Section~\ref{sec:problem-formulation}.
Furthermore, it is already proven to perform well in a similar scenario and problem tackled in~\cite{ozturk2018energy}.
In the following paragraphs, the proposed model for the employed $Q$-learning algorithm will be elaborated.

\subsection{Actions and States}
In the considered scenario, the IoT devices are supposed to choose one of the options presented in Fig.~\ref{fig:iot-options} to conduct their connection and data processing tasks.
In this regard, these options could also be treated as an action set for the developed $Q$-learning algorithm.
However, Options B, C, and E include either fog or cloud processing, meaning that IoT devices are supposed to offload their collected data to the fog or cloud for processing if they choose one of these options.
Provided that partial offloading is also captured in this work, the amount of data to be offloaded should also be optimized, and thus considering the options in Fig.~\ref{fig:iot-options} alone as the action set would not be adequate for this objective.
Therefore, the action set is determined as follows:
\begin{equation}\label{eq:action-set-first}
	\mathbb{A}=\mathbb{K} \times \Psi,
\end{equation}
where $\times$ represents a Cartesian product, $\mathbb{K}$ is the set of all the possible options included in Fig.~\ref{fig:iot-options}, such that $\mathbb{K} = \{A, B, C, D, E\}$, and $\Psi$ is the set of all the possible options for offloading percentage, such that
\begin{equation}
	\Psi = \{\tau m~|~m \in\{0,1,2,...,20\}\},\quad \psi\in \mathbb{R}^+,
\end{equation}
where $m$ is discretisation factor that is used to discretise the continuous values from $0\%$ to $100\%$, and $\tau$ is the resolution of the discretisation process.
$\tau$ can take any value in $\mathbb{R}^+$, but with a trade-off: the smaller it gets, the higher the resolution is, resulting in a more precise decision.
However, smaller $\tau$ creates an additional computational burden, since it increases  $|\mathbb{A}|$, which is the cardinality of $\mathbb{A}$.
Without loss of generality, $\tau=5$ is taken in this work, since it provides a sufficient resolution without significantly increasing the computational complexity.  
Note that Options A and D do not have $\Psi$ parameter, because they do not perform any offloading at all.
Using this phenomena, the actions set in \eqref{eq:action-set-first}, can be rewritten as
\begin{equation}
	\mathbb{A}= 
	\begin{dcases}
	\mathbb{K} \times \Psi,& \text{if } K\in \{\text{B},\text{C},\text{E}\}\\
	\mathbb{K},              & \text{if }  K \in \{\text{A},\text{D}\}.
	\end{dcases}
\end{equation} 

Since each of these actions also defines the state of the agent, the state space, denoted by $\mathbb{S}$ is designed to be the same with the action space, such that $\mathbb{S} = \mathbb{A}$.

\subsection{Penalty Function}
In the proposed $Q$-learning algorithm, two novel prioritization concepts are adopted:
	\paragraph{Prioritization of requirements} IoT devices are allowed to prioritize their requirements using a weighting mechanism, such that $w=\{w_\text{r}, w_\text{s}\}$, where $w_\text{r}\in\mathbb{R}$ and $w_\text{s}\in\mathbb{R}$ are the weight parameters for response time and security requirements, respectively.
	IoT devices are asked to rate the strictness of their requirements, such that lower values indicate that the requirement is loose, while higher values yield a stricter requirement.

	\paragraph{Prioritization of energy consumption and monetary cost} the total energy consumption and monetary cost are subject to prioritization as well.
	However, unlike the requirement prioritization case, where IoT devices control their weights~($w_\text{r}$ and $w_\text{s}$), the energy consumption and cost prioritization are triggered by the network.
	Moreover, this mechanism is linked to the battery level of an IoT device, $\beta$.
	More specifically, a certain threshold, denoted by $\beta_T \in \mathbb{R}$, is determined for the battery level, and 
	\begin{itemize}
		\item if the battery level of an IoT device $i$ is above or equal to the threshold, such that
		\begin{equation}
			\beta_i \geq  \beta_T,
		\end{equation}
		the monetary cost is prioritized.
		\item if, on the other hand, the battery level of an IoT device $i$ is less than the threshold, such that
		\begin{equation}
			\beta_i < \beta_T,
		\end{equation}
		then the energy consumption is prioritized.
	\end{itemize} 

Based on that, the overall penalty function for the developed $Q$-learning algorithm is formulated as follows:
\begin{equation}\label{eq:iot-penalty}
	\mathfrak{C}_Q = \Theta_\text{r} + \Theta_\text{s}  + \Theta_\text{c} + \Theta_\text{m} + w_\text{e} \hat{E}_\mathbb{T},
\end{equation}
where $\Theta_\text{r}$, $\Theta_\text{s}$, $\Theta_\text{c}$, and $\Theta_\text{m}$ are the penalty elements for response time, security, computational capacity, and monetary cost, respectively:
\begin{subequations}
	\begin{eqnarray}
	\Theta_\text{r}&=&
	\begin{dcases}\label{eq:response-time-cost}
	\Omega^{w_\text{r}}+\hat{K}_\text{r},& \text{if } \hat{K}_\text{r} > K_\text{r}\\
	0,               &\text{otherwise},
	\end{dcases}\\
	\Theta_\text{s}&=& 
	\begin{dcases}\label{eq:security-cost}
	\Omega^{w_\text{s}}, &\text{if } \hat{K}_\text{s} < K_\text{s}\\
	0,             &  \text{otherwise},
	\end{dcases} \\
	\Theta_\text{c}&=& 
	\begin{dcases}
	\Omega^{w_\text{c}}, &\text{if } K_\text{x} > \hat{K}_\text{X}\\
	0,            &   \text{otherwise},
	\end{dcases}\\\label{eq:monetary-cost}
	\Theta_\text{m}&=& w_\text{m}M_{\mathbb{T},S}\psi K_\text{x},
	\end{eqnarray}
\end{subequations}
where $\Omega \in \mathbb{R}$ is the global penalty factor, and $w_\text{c}\in \mathbb{R}$ is the penalty factor incurred when the computational capacity is exceeded.
Note that $w_\text{c}>\max\{w_\text{r},w_\text{s}\}$, since the computational capacity is a physical constraint that cannot be breached.

$w_\text{e}$ and $w_\text{m}$ are the weights for energy consumption and monetary cost set by the network, such that
\begin{subequations}
	\begin{eqnarray}
	w_\text{e}&=&
	\begin{dcases}\label{eq:energy-weight}
	w_\text{e}^+,& \text{if } \beta_i < \beta_T\\
	w_\text{e}^-,               &\text{otherwise},
	\end{dcases}\\
	w_\text{m}&=& 
	\begin{dcases}\label{eq:monetary-weight}
	w_\text{m}^+, &\text{if } \beta_i \geq  \beta_T\\
	w_\text{m}^-,             &  \text{otherwise},
	\end{dcases}
	\end{eqnarray}
\end{subequations}
where $x^+$ and $x^-$ represents the high and low values of $x$, respectively.

\section{Performance Evaluation}\label{sec:performance}
In this section, the proposed RL approach is implemented in a simulation environment, as illustrated in Fig.~\ref{fig:simuli}, using the simulation parameters defined in Table~\ref{tab:sim}.
\begin{table}[]
\centering
	\caption{Simulation parameters}
	\label{tab:sim}
	\begin{tabular}{@{}lll@{}}
		\toprule
		\multicolumn{1}{l}{\textbf{Parameter}} & \multicolumn{1}{l}{\textbf{Value}} & \multicolumn{1}{l}{\textbf{Description}}                 \\ \midrule
		\multicolumn{3}{c}{\textbf{Communication}}                                                                                             \\ \\
		$\delta$                            & $3$                                & Path loss exponent                                       \\
		$d_0$                                  & $10$ m                             & Reference distance                                       \\
		$\sigma$                               & $8$ dB                             & Standard deviation \\
		$c$                                    & $3\times10^8$ m/s                  & Speed of light                                           \\
		$f_\text{c,a}$                         & $2.4$ GHz                          & Carrier frequency for IEEE 802.11g                       \\
		$f_\text{c,b}$                         & $1700$ MHz                         & Carrier frequency for NB-IoT                             \\
		$f_\text{c,c}$                         & $1800$ MHz                         & Carrier frequency for LTE                                \\
		$\mathcal{N}_0$                                  & $-204$ dbW/Hz                      & Noise density                                            \\
		$T$                                    & $1/N$ s                            & Time period                                              \\
		$B$                                    & $180$ kHz                          & Bandwidth                                                \\
		$F$                                    & $2$                                & IEEE 802.11g retransmission rate\\ 
		$r_\text{enb}$ & $1$ km & Coverage radius of eNB\\
		$r_\text{wifi}$ & $30$ m & Coverage radius of eNB\\ \\
		\multicolumn{3}{c}{\textbf{General}}                                                                                                   \\ \\
		$N_\text{iot}$                                    & $10$                               & Number of IoT devices                                    \\
		$\epsilon$                             & $5\times10^{-6}$ J                 & Energy per computation cycle               \\
		$W$                                    & $8$                                & Number of bits per DE\\
		$\chi_\text{d}$                        & $30$ kbps                          & Computational capacity of device                         \\
		$\chi_\text{f}$                        & $100$ kbps                         & Computational capacity of fog                            \\
		$\chi_\text{c}$                        & $10$ Mbps                          & Computational capacity of cloud                          \\
		$X_\text{d}$                        & $100$                              & (Device) Comp. cycles per DE  \\
		$X_\text{f}$                        & $10$                               & (Fog) Comp. cycles per DE     \\
		$X_\text{c}$                        & $1$                                & (Cloud) Comp. cycles per DE   \\
		$M_\text{d}$                           & $10^{-4}$ AC                    & (Device) Cost of processing per bps                      \\
		$M_\text{f}$                           & $10^{-1}$ AC                    & (Fog) Cost of processing per bps                         \\
		$M_\text{c}$                           & $1$ AC                            & (Cloud) Cost of processing per bps                       \\ 
		$\mho$								   & $200$ & Data compression rate\\
		$\beta_T$ & $30\%$ & Threshold for battery level\\ \\
		\multicolumn{3}{c}{\textbf{$Q$-learning}}                                                                                              \\ \\
		$\alpha$                               & $0.5$                              & Learning rate                                            \\
		$\varphi$                                 & $0.9$                              & Discount factor                                          \\
		$\varepsilon$ 						   & $0.8$ 	& Chance of choosing random action\\
		$N_\text{ep}$                          & $10^3$                             & Number of episodes                                       \\
		$N_\text{it}$                          & $10^3$                             & Number of iterations per episode\\
		$\Omega$ & $10$ &  Global penalty factor\\
		$w_\text{c}$ & $5$ & Penalty of exceeding comp. capacity \\
		$w_\text{e}^+$, $w_\text{e}^-$ & $10$, $1$ & High, low values of $w_\text{e}$\\
		$w_\text{m}^+$, $w_\text{m}^-$ & $10$, $0$ & High, low values of $w_\text{m}$                        \\ \bottomrule
	\end{tabular}
\end{table}

\subsection{Benchmarking}
In order to obtain the benchmark scenarios, first, the IoT devices are categorized into two groups---with equal number of members---based on their wireless connection type as follows: the devices with NB-IoT connection and the ones with Wi-Fi connection.
The former group is called as Group-X, while the latter group represents Group-Y.
Then, these groups are mapped to the available processing units, and all the possible combinations are considered as benchmark scenarios, as given in Table~\ref{tab:iot-benhmarks}.

\begin{table}[h]
	\centering
	\caption{List of fixed benchmark scenarios with connection types and data processing unit}
	\label{tab:iot-benhmarks}
	\begin{tabular}{@{}ccc@{}}
		\toprule
		\textbf{Scenario} & \textbf{Group-X} & \textbf{Group-Y} \\ \midrule
		Sc$_\text{A}$                 & Device           & Device           \\
		Sc$_\text{B}$                 & Cloud            & Device           \\
		Sc$_\text{C}$                & Device           & Fog              \\
		Sc$_\text{D}$                 & Cloud            & Fog              \\
		Sc$_\text{E}$                 & Device           & Cloud            \\
		Sc$_\text{F}$                 & Cloud            & Cloud            \\ \bottomrule
	\end{tabular}
\end{table}

\subsection{Performance Metrics}
The obtained results are evaluated in five different metrics, namely energy consumption, monetary cost, response time, security dissatisfaction, and a novel joint metric that is specifically developed for this work.
Moreover, these metrics are presented in a comparative fashion, where the performances of the benchmark methods---provided in Table~\ref{tab:iot-benhmarks}---are compared to the proposed RL-based method.

The performance metrics can be elaborated as follows:
	\paragraph{Energy consumption} the accumulated energy consumption of all the IoT devices, which is caused by data processing and transmission, is calculated by
	\begin{equation}
		\dot{E}_\mathbb{T}=\sum_{i=1}^{N_\text{iot}}E_{\mathbb{T},i},
	\end{equation}
	where $N_\text{iot}$ is the number of IoT devices, and $E_{\mathbb{T},i}$ is the total energy consumption of $i^\text{th}$ IoT device.
	
	\paragraph{Monetary cost} the aggregated monetary cost that the IoT devices are charged for data processing:
	\begin{equation}
	\dot{M}_\mathbb{T}=\sum_{i=1}^{N_\text{iot}}M_{\mathbb{T},i},
	\end{equation}
	where $M_{\mathbb{T},i}$ is the total monetary cost for $i^\text{th}$ IoT device.
	
	\paragraph{Response time} the accumulated response time that occurs for all the IoT devices:
	\begin{equation}
	\dot{R}_\mathbb{T}=\sum_{i=1}^{N_\text{iot}}R_{\mathbb{T},i},
	\end{equation}
	where $R_{\mathbb{T},i}$ is the total response time occurred with $i^\text{th}$ IoT device, and given by
	\begin{equation}\label{eq:res-time-condition}
		R_{\mathbb{T},i} = 
		\begin{dcases}
		\hat{K}_\text{r} - K_\text{r}, &\text{if } \hat{K}_\text{r} > K_\text{r}\\
		0,             &  \text{otherwise.}
		\end{dcases}
	\end{equation}
	
	\paragraph{Security dissatisfaction} the number of IoT devices, whose security requirements are not satisfied:
	\begin{equation}
		\dot{N}_{\text{dis},\mathbb{T}}=\sum_{i=1}^{N_\text{iot}}v_{\text{dis},i},
	\end{equation}
	where $v_{\text{dis},i}$ is the security dissatisfaction variable for the IoT device $i$, such that
	\begin{equation}
		v_\text{dis}=
		\begin{dcases}
	     0, &\text{if } 	\hat{K}_\text{s} \geq K_\text{s}\\
	     1,             &  \text{otherwise}.
		\end{dcases}
	\end{equation}
	
	\paragraph{Joint metric} the combination of all the aforementioned metrics, such that
	\begin{equation} \label{eq:joint-metric}
		J = \gamma\hat{\ddot{\mathsf{E}}}_\mathbb{T} + \eta\hat{\ddot{M}}_\mathbb{T} + \zeta\hat{\ddot{R}}_\mathbb{T} + \kappa\hat{\ddot{N}}_{\text{dis},\mathbb{T}},
	\end{equation}
	where $\hat{\ddot{\mathsf{E}}}_\mathbb{T}$, $\hat{\ddot{M}}_\mathbb{T}$, $\hat{\ddot{R}}_\mathbb{T}$, and $\hat{\ddot{N}}_{\text{dis},\mathbb{T}}$ are the normalised versions of $\ddot{\mathsf{E}}_\mathbb{T}$, $\ddot{M}_\mathbb{T}$, $\ddot{R}_\mathbb{T}$, and $\ddot{N}_{\text{dis},\mathbb{T}}$, respectively.
	The normalisation~(feature scaling) operation is performed here in order to keep the scale of the each metric in the same range, thus preventing one from dominating another.
	$\gamma$~(unitless), $\eta$ in (Joule/AC), $\zeta$ in (Joule/s), and $\kappa$ in (Joule), where $\gamma=\eta=\zeta=\kappa=1$, are coefficients used to make the units of the elements of $J$ in \eqref{eq:joint-metric} the same.
	Note that AC in the unit of $\eta$ stands for arbitrary currency.

\subsection{Battery Regimes}
Two different battery regimes, namely low and high, are considered in this work, and results are produced separately in order to observe the behaviours of the proposed method\footnote{It is worth noting here that the benchmark methods do not consider the remaining battery level of IoT devices, and thus their behaviours are not expected to change with the battery level.}.
	\paragraph{Low-battery regime} when the energy level of the battery of a particular IoT device is under a certain threshold~($\beta_i < \beta_T$), the device is treated as in a low-battery regime, and the proposed $Q$-learning algorithm starts to prioritize the total energy consumption of the device along with meeting the response time and security requirements.
	It is worth noting that, the priority between the requirements and the energy consumption is determined by the weights of the requirements~($w$), such that if the requirements are strict~(i.e., with high weights), then they become more important than the energy consumption, and vice versa.
	The underlying idea here is that if the device is strict in any requirement, it means that it is unwilling to compromise on that.
	Moreover, the monetary cost is completely discarded, and therefore IoT devices are expected to be charged more when they are in this regime.
	
	\paragraph{High-battery regime} to be in the high-battery regime, the remaining energy in the battery of an IoT device should be above the aforementioned threshold for the battery level~($\beta_i \geq \beta_T$).
	In this high-battery regime, the monetary cost is valued significantly, and---similar to the low-battery regime---the importance of the requirements are determined by their correspondent weights, and energy consumption is loosely prioritized\footnote{It is important to mention that unlike the low-energy regime, where the monetary cost is completely discarded, the energy consumption is still considered in the high-battery regime albeit with much less importance, since energy consumption of an IoT device should always be in the equation.}.

\subsection{Results and Discussions}
\begin{figure}[h]
	\centering
	\includegraphics[width=\columnwidth]{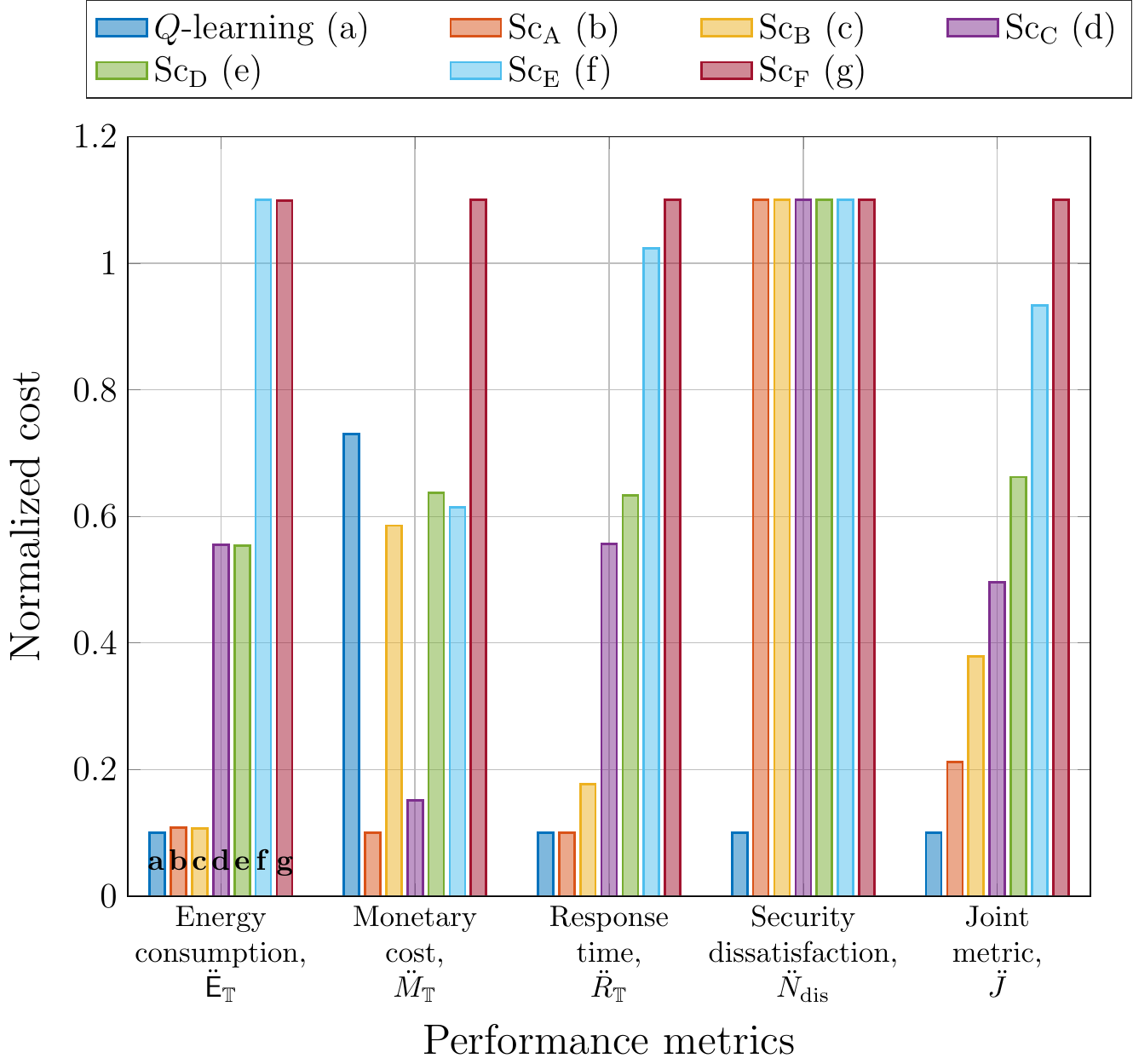}
	\caption{Performances of the proposed method~(named as $Q$-learning in the legend) and the benchmark scenarios in terms of considered metrics when all the IoT devices are in the low-battery regime. The response time and security requirements are strictly prioritized, such that $w_\text{r}=w_\text{s}=3$. Note that the shown results are normalised values in the range of $[0,1]$ with an offset of $0.1$, which is used only due to visualisation purposes. However, the results are discussed in the text without considering the offset value. The indexing from (a) to (g) in the legend and the first set of bars are done for identification, and the same order follows for all sets of the bars.}
	\label{fig:low_battery-res_time_3-sec_3}
\end{figure}

Fig.~\ref{fig:low_battery-res_time_3-sec_3} demonstrates the performances of all the methods including the proposed one and the benchmarks when the entire set of IoT devices are in the low-battery regime.
Moreover, both response time and security requirements of the IoT devices are prioritized in a rigid way, where both $w_\text{r}$ and $w_\text{s}$ are ranked as $3$.
In the following paragraphs, there will be individual discussions on the results for each performance metric:
	\paragraph{Energy consumption} owing to the low-battery regime, it was expected that the proposed method will perform well in minimizing the energy consumption.
	This is due to the fact that, when the battery level is under the threshold, the energy consumption component in the penalty function in \eqref{eq:iot-penalty} is prioritized through its weight~($w_\text{e}$) by setting it to its high value~($w_\text{e}^+$), as seen in \eqref{eq:energy-weight}.
	This is a reasonable behaviour, since the energy consumption becomes more crucial when the battery is about be depleted, which in turn interrupts the communication until the battery is recharged or replaced.
	
	The energy consumption results for the benchmark scenarios are also worth discussing.
	There are two main components of the overall energy consumption, namely data processing and transmission.
	Based on the channel conditions and distance between a transmitter and a receiver, the transmission energy consumption can prevail over the processing energy consumption, or vice versa~\cite{TayadeOct17}.
	However, receiver sensitivity, which is captured by the link margin in this work, also plays an important role provided that it is directly correlated to the required received power, which in turn affects the required transmit power.
	Therefore, the connection type~(i.e., Wi-Fi, NB-IoT, and LTE) is also involved in the breakdown of the total energy consumption.
	In this work, due to the random distribution of the IoT devices and the Wi-Fi gateways at each repeat\footnote{The simulations are repeated for 25 times to avoid random effects.} in the simulations, it is avoided for one of the aforementioned two components of energy consumption to dominate the other\footnote{This is ensured via random process; the path-loss can sometimes be huge, which results in the transmission energy consumption surpassing the data processing energy consumption, or vice versa. However, this effect is minimized by averaging out the simulation repeats.}. 
	Nonetheless, owing to the link margin assumption, the NB-IoT connection happened to be the least energy consuming connection type in most of the cases.
	Besides, in terms of the energy consumption, the descending order of the tasks is as follows: Wi-Fi connection, data processing, NB-IoT connection. 
	Note that the energy consumption difference between the Wi-Fi connection and data processing happened to be much more than the difference between the data processing and NB-IoT connection.
	This may seem counter-intuitive given that the Wi-Fi gateway is much closer to the IoT devices than the eNB.
	Nonetheless, since the receiver sensitivity of NB-IoT is less than Wi-Fi and LTE, it happens to result in less energy consumption owing to the less required transmit power, which is caused by the less path-loss.
	
	In this regard, since all the options include the same number of IoT devices with NB-IoT~(Group-X) and Wi-Fi~(Group-Y), there is no difference in terms of the number of connection types.
	However, the point that matters here is the processing unit.
	On one hand, when an IoT devices is connected through Wi-Fi, device processing is expected to consume less energy than cloud processing, with fog processing in between.
	On the other hand, when the device is connected through NB-IoT, device processing consumes more energy than cloud processing, with fog processing in between.
	In addition, the maximum energy consumption with Wi-Fi connection is expected to be more than the maximum energy consumption with NB-IoT connection.
	Based on that, Sc$_\text{B}$ resulted in the least energy consumption due to the fact that the devices with Wi-Fi connection process the data locally, while the devices using NB-IoT connection performs cloud processing.
	Sc$_\text{B}$ is followed by Sc$_\text{A}$, since it also processes the data locally for Wi-Fi connections.
	The small difference between Sc$_\text{A}$ and Sc$_\text{B}$ comes from the difference between the cloud and device processing for NB-IoT connections.
	In a similar fashion, Sc$_\text{E}$ consumed the maximum energy among the benchmark methods, since it employs cloud processing for Wi-Fi connected devices and device processing for NB-IoT connected devices.
	In summary, cloud processing is less energy consuming for NB-IoT, whereas it is more energy consuming for Wi-Fi.
	Moreover, as mentioned earlier, the maximum energy consumption with Wi-Fi is much more than that of NB-IoT.
	
	\paragraph{Monetary cost} as a result of the low-battery regime, the proposed algorithm inclines towards being much looser in monetary cost, and thus it is not expected to be competitive in this metric.
	Based on that, as expected, the proposed method performed worse than all the benchmark methods other than Sc$_\text{F}$, which purely includes cloud processing.
	
	Similarly, the results of the benchmark methods are obtained as expected: the cloud is the most expensive means of data processing, followed by the fog, and the device~(local), respectively, such that $M_\text{d}<M_\text{f}<M_\text{c}$.
	Thus, the scenarios with device processing~(e.g., Sc$_\text{A}$) resulted in less amount of monetary cost, whereas the scenarios with cloud and/or fog processing~(e.g., Sc$_\text{D}$ and Sc$_\text{F}$) become the most expensive ones.
	The results obtained in Fig.~\ref{fig:low_battery-res_time_3-sec_3} confirms this statement.
	
	\paragraph{Response time} the proposed method performed quite well in response time by outperforming all the benchmark methods.
	Provided that $w_\text{r}=3$, which yields a strict prioritization of response time, it was expected that the proposed method will reduce the response time.
	Considering \eqref{eq:iot-penalty} and \eqref{eq:response-time-cost} together, the effect of response time in the penalty function in~\eqref{eq:iot-penalty} increases with growing $w_\text{r}$. 
	As such, the primary objective of the developed $Q$-learning algorithm is to minimize the overall penalty, hence, the response time satisfaction becomes key for this objective.
	
	Similar to the previous metrics, the benchmark methods also performed as anticipated.
	As discussed in Section~\ref{sec:response-time-model}, response time is a function of the number of hops~($N_\text{h}$), the computational power~($X$), retransmission rate~($F$), the data volume~($D\in\{D_\text{r},D_\text{p}\}$), and compression rate~($\mho$).
	Given that NB-IoT connection has only one hop and lower retransmission rate than Wi-Fi connection, the scenarios with NB-IoT resulted in a comparatively less response time.
	Similarly, from~\eqref{eq:response-time-model}, albeit suffering from a higher computational time, device processing is also preferable due to the processed data transmission, which entails $\mho$ times less data volume\footnote{The response time of the options~(connection and processing) would alter for different values of $\mho$, $F$, and $X$. Therefore, the response time of the benchmark methods would change accordingly, but the discussions here are based on the current assumptions for $\mho$, $F$, and $X$.}.
	Thus, for example, a cloud processing with a Wi-Fi connection would result in the highest response time owing to: 1) Wi-Fi connection, which has higher retransmission rate; 2) two hops taken; and 3) raw data transmission.
	In this regard, Sc$_\text{A}$ resulted in the least response time, whereas Sc$_\text{F}$ caused the highest response time among all the methods.
	
	\paragraph{Security dissatisfaction} similar to the response time case, security requirement is also strictly prioritized in these simulation campaigns by setting $w_\text{s}$ to $3$.
	Considering \eqref{eq:iot-penalty} with \eqref{eq:security-cost}, higher values of $w_\text{s}$ incurs more cost by increasing $\Theta_\text{s}$, which in turn inflates the penalty function, $\mathfrak{C}_Q$ in~\eqref{eq:iot-penalty}.
	Given that the objective of the designed $Q$-learning algorithm is to minimize $\mathfrak{C}_Q$, satisfying the security requirement of IoT devices becomes crucial for the proposed algorithm, as $\Theta_\text{s}$ returns 0 when the requirement is met.
	To this end, the developed $Q$-learning algorithm achieved a significant reduction in terms of the security dissatisfaction when compared to the benchmark methods.
	
	One can question the equal results of the benchmark methods, but there is a rationale behind it: half of the IoT devices are connected with NB-IoT~(Group-X in Table~\ref{tab:iot-benhmarks}), while the other half communicates through Wi-Fi~(Group-Y in Table~\ref{tab:iot-benhmarks}), and---as discussed in Section~\ref{sec:problem-formulation}---the security requirement is captured by the need for an eSIM card, which is only available for NB-IoT connections.
	Thus, those connected with NB-IoT do not have any issue with the security dissatisfaction, since they always meet the requirements due to their eSIM card availability.
	Those connected through Wi-Fi, on the other hand, cannot respond to the eSIM card requirement.
	Based on that, the number of IoT devices with security dissatisfaction always equals to the number of IoT devices that: 1) is connected through Wi-Fi and 2) requires eSIM protection.
	Thus, the number of dissatisfied devices is the same for all the benchmark methods.
	
	\paragraph{Joint metric} while each individual previous metric reflects the behaviours of the methods in a specialized manner, this joint metric summarizes the overall performances.
	Therefore, this metric can be seen as a holistic cost of each method, which demonstrates how they performed when all the previous metrics are combined.
	The proposed method performed quite well and outperformed all the benchmark methods in different scales.
	The reasoning behind this is that there are four metrics in total other than joint metric, and the proposed method outperformed all the benchmark methods in three of them.
	Therefore, it is quite reasonable that the proposed method performed the best in terms of the joint metric.
	Similarly, the benchmark methods responded to the joint metric according to their results in each individual metric, namely: energy consumption, monetary cost, response time, and security dissatisfaction. 

\begin{figure}[h]
	\centering
	\includegraphics[width=\columnwidth]{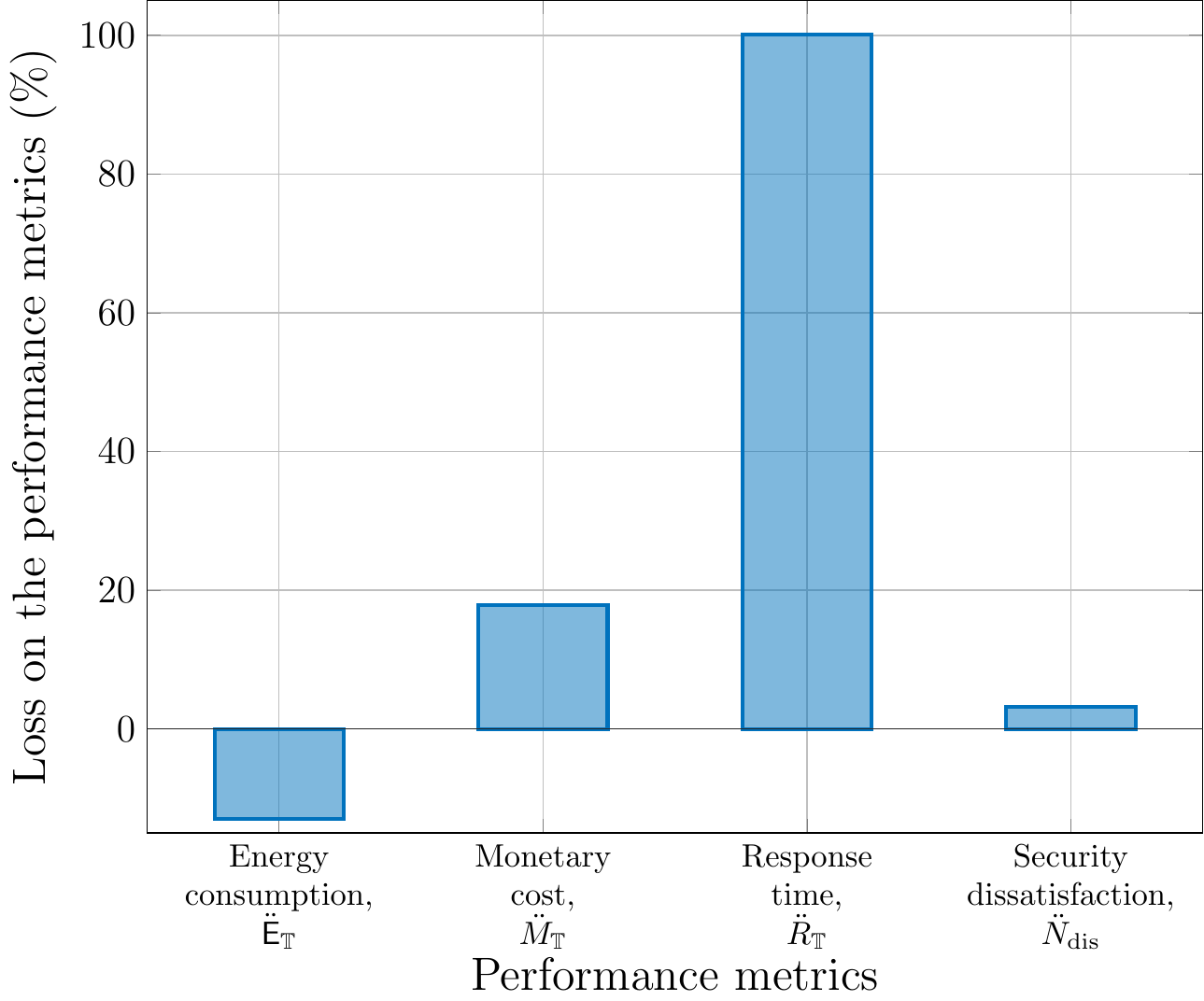}
	\caption{Performances of the proposed method in terms of considered metrics when all the IoT devices are in the low-battery regime. The results show the percentage loss when LBRU~($w_\text{r}=w_\text{s}=0$) is compared to LBRA~($w_\text{r}=w_\text{s}=3$). Positive values yield loss; i.e., the superiority of LBRA, while negative values yield gain; i.e., the superiority of LBRU.}\label{fig:low_battery-comparison}
\end{figure}

The comparison between the cases of requirement de-prioritization~($w_\text{r}=w_\text{s}=0$)---which will be referred to as low-battery requirement-aware~(LBRA) hereafter---and prioritization~($w_\text{r}=w_\text{s}=3$)---which will be referred to as low-battery requirement-unaware~(LBRU) hereafter---under the low-battery regime is demonstrated in Fig.~\ref{fig:low_battery-comparison}.
The loss calculations for each metric is performed as follows:
\begin{equation}\label{eq:iot-gain}
L = \dfrac{\Delta_{Q,\text{LBRU}}-\Delta_{Q,\text{LBRA}}}{\Delta_{Q,\text{LBRU}}},
\end{equation}
where $\Delta_{Q,\text{LBRA}}$ and $\Delta_{Q,\text{LBRU}}$ are the differences between the values obtained via the proposed method and the minimum value obtained with the benchmark methods for LBRA and LBRU, respectively, such that
\begin{subequations}
	\begin{eqnarray}
		\Delta_{Q,\text{LBRA}} = V_{Q,\text{LBRA}} - \min(V_\text{b}),\\
		\Delta_{Q,\text{LBRU}} = V_{Q,\text{LBRU}} - \min(V_\text{b}),
	\end{eqnarray}
\end{subequations}
where $V_\text{b} \in \{V_{Sc_\text{A}}, V_{Sc_\text{B}}, V_{Sc_\text{C}}, V_{Sc_\text{D}}, V_{Sc_\text{E}}, V_{Sc_\text{F}} \}$ is the obtained values in the aforementioned metrics with the benchmark methods.
$ V_{Q,\text{LBRA}}$ and $ V_{Q,\text{LBRU}}$ are the obtained values in the aforementioned metrics with the proposed method for LBRA and LBRU, respectively.

There is an important caveat to be noted here: positive values of $L$ calculated through~\eqref{eq:iot-gain} indicate loss, where the LBRU performed worse than LBRA, and the negative values of $L$ indicate gain, where LBRU performed better than LBRA.
From the findings in Fig.~\ref{fig:low_battery-comparison}, it is obvious that LBRA outperformed LBRU in all the metrics other than the energy consumption.
Starting from response time and security dissatisfaction, the results are quite expected provided that both of the requirements are prioritized in LBRA with weight values of $w_\text{r}=w_\text{s}=3$.
Hence they are given a special care in LBRA when~\eqref{eq:iot-penalty} is considered together with \eqref{eq:response-time-cost} and \eqref{eq:security-cost}.
Although the scale difference between the response time and security dissatisfaction is worth discussing, however, it is better to first analyze the energy consumption results, which will then be more beneficial to explain such difference.

As seen from Fig.~\ref{fig:low_battery-comparison}, the energy consumption is the only metric that LBRU performed better than LBRA.
The rationale behind is that energy consumption is the only focus of LBRU given that: 1) the IoT devices are in the low-battery regime, and 2) both $w_\text{r}$ and $w_\text{s}$ are set to 0.
Thus, LBRU does not consider any other metric other than energy consumption, which is the reason why it managed to outperform LBRA.

Considering this rationale as a base, it would be more straightforward to explain the obtained results from the other metrics, as the reasoning for all the results are linked to each other and they all arise from this base.
There are additional supporting facts as follows:
\begin{itemize}
	\item LBRA aims at minimizing the energy consumption, but with the response time and security constraints.
	In other words, the energy consumption is minimized after the response time and security requirements of the devices are satisfied.
	Provided that Options A, D, and E happen to result in minimal response time, meaning that they could be the options for the devices with low response time requirements.
	\item similarly, Options D and E are the ones that provide the eSIM protection, which means an IoT device should select one of these if it has security concerns.
\end{itemize}
Based on that, Options D and E followed by A are the intersection ones that are most likely to be selected when both response time and security is prioritized.
For example, the response time requirement of an IoT device can only be satisfied with Option A and D, but Option E can result in less energy consumption.
In such cases, LBRA would select Option A or D that has the least monetary cost due to device processing, whereas LBRU goes for Option E that result in the highest monetary cost due to cloud processing.
As such, LBRA is more likely to perform better in terms of monetary cost, while LBRU is better in energy consumption.
These explain the performance differences between LBRA and LBRU in terms of energy consumption and monetary cost.

It is now better to turn back to the discussion on the scale difference between the response time and security dissatisfaction.
There are two points to consider:
\begin{itemize}
	\item switching among Wi-Fi options~(i.e., A, B, and C) and among NB-IoT options~(i.e., D and E) do not change the security dissatisfaction results, but it changes the response time.
	In other words, selecting a different option from Fig.~\ref{fig:iot-options} definitely changes the response time behaviour, whereas the security behaviour might remain the same.
	In this regard, there is more room for response time to alter than that of the security behaviour.
	
	\item Options D and E have NB-IoT connection, which was already discussed in this section as being the least energy consuming one in majority of the cases, and thus the agent would be more prone to stick with them.
	Owing to the fact that Options D and E are with NB-IoT connection, both LBRA and LBRU would more possibly select one of these options, which would have an impact on the response time but the security behaviour remains unaffected.
\end{itemize}
Due to these reasons, it is quite reasonable that LBRA outperforms LBRU more significantly in response time than that in security dissatisfaction.

Table~\ref{tab:high-low_battery-comparison} reveals the performance comparison between LBRA and the case when the IoT devices are in the high-battery regime and their requirements are fully prioritized~($w_\text{r}=w_\text{s}=3$)---which will be referred to as high-battery requirement-aware~(HBRA) hereafter.
The results show the percentage loss when HBRA is compared to LBRA.
Positive values yield loss; i.e., the superiority of LBRA, while negative values yield gain; i.e., the superiority of HBRA.
Note that only energy consumption and monetary cost results are presented in Table~\ref{tab:high-low_battery-comparison}, since LBRA and HBRA performed equally well in response time and security dissatisfaction given that they both fully prioritize the device requirements.
\begin{table}[]
\centering
\caption{Loss on the given performance metrics}
\label{tab:high-low_battery-comparison}
\begin{tabular}{@{}cc@{}}
\toprule
\textbf{\begin{tabular}[c]{@{}c@{}}Energy consumption\\ $\ddot{\mathsf{E}}_\mathbb{T}$\end{tabular}} & \textbf{\begin{tabular}[c]{@{}c@{}}Monetary cost\\ $\ddot{M}_\mathbb{T}$\end{tabular}} \\ \midrule
54.1525\%                                                                                            & -96.7937\%                                                                             \\ \bottomrule
\end{tabular}
\end{table}

The findings in Table~\ref{tab:high-low_battery-comparison} show that LBRA reduced the energy consumption of HBRA by around $53\%$, while HBRA managed to decrease the monetary cost of LBRA by around $97\%$.
These results are quite expected because LBRA focuses only on the energy consumption, while completely ignoring the monetary cost reduction\footnote{Energy consumption minimization and/or monetary cost reduction are the secondary objectives for both LBRA and HBRA, since they fully prioritize the device requirements.}.
HBRA, on the other hand, aims at minimizing the monetary cost rather than the energy consumption, since the battery levels of IoT devices are high.

Although these explanations are adequate to understand why they surpass each other in the two considered metrics, there is still room for clarification for the question of why the scales of the outperformance are quite different from each other.
Considering \eqref{eq:iot-penalty} together with \eqref{eq:monetary-cost}, \eqref{eq:monetary-weight}, and \eqref{eq:energy-weight}, it is obvious that both energy consumption and monetary cost have their own impacts in $\mathfrak{C}_Q$.
On one hand, when the IoT devices are in the low-battery regime, the weight for energy consumption~($w_\text{e}$) takes its higher value~($w^+_\text{e}$), which is set to 10, while the weight for monetary cost~($w_\text{m}$) takes its lower value~($w^-_\text{m}$), which is set to 0.
On the other hand, when the IoT devices are in the high-battery regime, $w_\text{e}$ is set to its lower value as $w^-_\text{e}=1$, while $w_\text{m}$ is set to its higher value as $w^-_\text{e}=10$.
This means that
\begin{itemize}
	\item when the IoT devices are in the low-battery regime, the algorithm fully focuses on the energy consumption minimization while completely discarding the monetary cost reduction; but
	\item when the IoT devices are in the high-battery regime, the algorithm mainly takes care of the monetary cost reduction, but without completely ignoring the energy consumption minimization.
\end{itemize}
This is done because energy consumption is always important for an IoT device regardless of the battery level, which would only change the degree of importance.
As such, HBRA still tries to conserve some energy while focusing primarily on the monetary cost reduction, and thus this puts a barrier for HBRA's loss in energy consumption.
Nonetheless, LBRA does not take the monetary cost reduction into account at all, as a result, the scale of HBRA's gain in monetary cost is more than HBRA's loss in energy consumption.

\section{Conclusion}\label{sec:conclusion}
A novel context-aware approach is presented in this work for IoT networks, and connectivity-processor pair is jointly optimized in order to meet the objectives in terms of the energy consumption, monetary cost, security, and response time.
More specifically, this work is an attempt to determine the wireless connection type and data processing unit along with the amount of data to be offloaded, which is the case when data to be processed at a unit other than the device.
In that regard, IoT devices come with diverse requirements in terms of response time and security, and they are allowed to prioritize their requirements.
In addition, the proposed scheme also takes the battery level of a device into account, such that the minimization of energy consumption becomes the focus if the battery level is under a certain threshold, while the reduction of monetary cost is mainly targeted in case the battery level is above that threshold.
The proposed scheme employs $Q$-learning algorithm, and manages to achieve significant gains compared to deterministic benchmark routes.
Results demonstrate that the  proposed method outperforms all the benchmark methods in the novel joint metric, combining all the objectives of the  formulated problem.

\bibliographystyle{IEEEtran}
\bibliography{ref}
\end{document}